\documentclass[lettersize,journal]{IEEEtran}
\usepackage{amsmath,amsfonts}
\usepackage{algpseudocode}
\usepackage{algorithm}
\usepackage{times}
\usepackage{array}
\usepackage[caption=false,font=normalsize,labelfont=sf,textfont=sf]{subfig}
\usepackage{textcomp}
\usepackage{stfloats}
\usepackage{url}
\usepackage{verbatim}
\usepackage{graphicx}
\usepackage{cite}
\usepackage{pifont}
\usepackage{graphicx} 
\usepackage{array}
\usepackage{amsmath}
\usepackage{amssymb}
\usepackage{xcolor}

\begin{document}

\title{FaaSMT: Lightweight Serverless Framework for Intrusion Detection Using Merkle Tree and Task Inlining}

\author{Chuang Li,~\IEEEmembership{Member,~IEEE,}
        Lanfang Huang,
        Gang Liu\textsuperscript{*},~\IEEEmembership{Member,~IEEE,}
        Dian He,
        Yanhua Wen,\\
        Lixin Duan,~\IEEEmembership{Member,~IEEE,}
        
\thanks{Manuscript created September 23, 2024,(* Corresponding author: Gang Liu)}
\thanks{Chuang Li, Lanfang Huang, Dian He, and Yanhua Wen are with the College of Computer Science, Hunan University of Technology and Business, and Xiangjiang Laboratory, Hunan 410205, China. (e-mail: chuangli@hutb.edu.cn, chouzhu152769614@gmail.com, hedian@hutb.edu.cn, yanhua-wen@hutb.edu.cn)}
\thanks{Gang Liu and Lixin Duan are with the Shenzhen Institute for Advanced Study, University of Electronic Science and Technology of China.(e-mail:liug@hnu.edu.cn, lxduan@uestc.edu.cn) }}

\markboth{Journal of \LaTeX\ Class Files,~Vol.~14, No.~8, October~2024}%
{Shell \MakeLowercase{\textit{et al.}}: A Sample Article Using IEEEtran.cls for IEEE Journals}

\maketitle

\begin{abstract}

The serverless platform aims to facilitate cloud applications' straightforward deployment, scaling, and management. Unfortunately, the distributed nature of serverless computing makes it difficult to port traditional security tools directly. The existing serverless solutions primarily identify potential threats or performance bottlenecks through post-analysis of modified operating system audit logs, detection of encrypted traffic offloading, or the collection of runtime metrics. However, these methods often prove inadequate for comprehensively detecting communication violations across functions. This limitation restricts the real-time log monitoring and validation capabilities in distributed environments while impeding the maintenance of minimal communication overhead. Therefore, this paper presents FaaSMT, which aims to fill this gap by addressing research questions related to security checks and the optimization of performance and costs in serverless applications. This framework employs parallel processing for the collection of distributed data logs, incorporating Merkle Tree algorithms and heuristic optimisation methods to achieve adaptive inline security task execution. The results of experimental trials demonstrate that FaaSMT is capable of effectively identifying major attack types (e.g., Denial of Wallet (DoW) and Business Logic attacks), thereby providing comprehensive monitoring and validation of function executions while significantly reducing performance overhead.

\end{abstract}


\section{Introduction}

\IEEEPARstart{S}{erverless} computing (or Functions-as-a-Service, FaaS) represents a contemporary architectural approach to the development of resilient and scalable applications \cite{alpernas2021cloud}. An increasing number of developers are opting to deploy applications in the cloud, as this model allows them to implement complex workflows through stateless, event-driven functions. These functions can be deployed on cloud platforms (e.g., AWS Lambda\footnote{\url{https://aws.amazon.com/lambda}}, Google Cloud Functions\footnote{\url{https://cloud.google.com/functions}}, and Microsoft Azure Functions\footnote{\url{https://azure.microsoft.com/products/functions/}}), and are widely used in scenarios including web services, API services, parallel processing, and machine learning pipelines \cite{carreira2019cirrus}. The cloud platform manages hardware, software runtimes, and operational tasks, enabling developers to focus on business logic while providing a flexible pay-as-you-go model \cite{daly2020event,kim2022vulnerabilities,li2022securing}. Unlike virtual machines, function instances are initiated only when there is a need to process requests, thereby improving resource utilization efficiency. Consequently, serverless computing has become a universal programming model for various applications \cite{fouladi2017encoding,jonas2017occupy,yan2016building}.

Despite the growing popularity of serverless computing, its distributed nature and reliance on third-party services and APIs present significant challenges \cite{li2022securing,wang2018peeking}. This dependency enhances development efficiency and resource utilization while simultaneously increasing security risks and complexity, thereby rendering application monitoring and detection more challenging. In particular, the deployment of each function as an independent microservice increases the potential attack surface \cite{chen2024microfi}, thereby enabling malicious actors to exploit potential vulnerabilities during network communication and data transfer between functions. For example, attackers may attempt to manipulate data in transit (e.g., data injection, tampering, or replay) to implement Denial of Wallet (DoW) strategies \cite{kelly2021denial,kuhlenkamp2020ifs,wang2018peeking}, thereby disrupting the normal operation of applications. Furthermore, the independent deployment of functions may reveal internal implementation details and interdependencies, which can be exploited by attackers to compromise execution paths and affect the correctness and performance of applications. Even minor modifications to critical execution attributes (e.g., call chains, memory usage, timeouts, and resource allocation) that functions depend on can result in deviations from the expected application behaviour.

To ensure the integrity of each task in a serverless application environment and optimize performance and cost-effectiveness through adaptive task fusion, several key challenges arise. Existing web application security tools, particularly log-based anomaly detection systems, have been adapted for serverless applications, primarily focusing on detecting known vulnerabilities and non-real-time attacks \cite{paccagnella2020custos,hassan2020omegalog}. Encouragingly, there are third-party tools (e.g., Kulium \cite{jegan2023guarding}) that offer real-time attack detection solutions with control flow protection and encrypted traffic interception techniques to enhance the data integrity of serverless applications. These tools mainly concentrate on system-level modifications, which can incur high-performance overhead, making them difficult to apply widely. In addition, to improve the performance optimization of serverless applications, tools like Fusionize \cite{schirmer2022fusionize} provide performance optimization and cost-benefit analysis functions that dynamically adjust optimization strategies based on load variations or application updates. Lin et al. \cite{lin2020modeling} have introduced a novel construct that formally defines serverless workflows and develops models for predicting average end-to-end response times and costs. Notably, their focus is primarily on performance and cost optimization, overlooking task verification. Therefore, to address the shortcomings of existing solutions regarding data integrity in serverless environments, we can adopt common serverless design patterns \cite{hong2018go,shafiei2022serverless,shafiei2019serverless,lannurien2023herofake} (e.g., statelessness, function orchestration, event-driven architecture, and input/output functions) to enhance the overall performance and reliability of the system.

Inspired by the above, we designed FaaSMT, an automated monitoring and analysis tool for detecting application violations. Our solution leverages the log collection mechanisms of FaaS platforms, combined with an efficient detection algorithm using Merkle Tree, to swiftly validate the overall trustworthiness of resource attributes for each function execution instance while ensuring remote FaaS function calls are extended through task source code. Specifically, we leverage the ubiquity of REST APIs to track the call chains of each task in event-driven serverless applications, analysing the trace log information using Merkle Tree to generate trust proofs. To achieve adaptive adjustments of application behaviour, FaaSMT employs a modified Continuous Sampling Plan (CSP-1) \cite{bermbach2011extendable,dodge1943sampling} to determine when to perform Merkle Tree verification on serverless applications to address suspicious activities, and it uses heuristic methods to identify the next optimal fusion configuration based on the verification results. We have implemented FaaSMT on the open-source serverless computing platform AWS and conducted a security assessment on two representative FaaS applications: ($1$) a commonly used trigger operation workflow in IoT environments and ($2$) TREE: a synthetic fan-out application. The results demonstrate that FaaSMT effectively examines function call information within workflows, thereby identifying potential security threats and enhancing the defence capabilities of serverless applications. This paper expands upon the work presented in \cite{schirmer2022fusionize} and makes the following key contributions:

• We propose FaaSMT, which integrates Merkle Tree algorithms and FaaS function optimization mechanisms to enable real-time monitoring and trust verification of serverless applications.

• FaaSMT tracks the task invocation chains in event-driven serverless applications, utilizing trace logs for behavioural analysis and generating trustworthy Merkle Tree proofs.

• FaaSMT can adaptively determine when to implement the verification mechanism and optimize fusion setup based on verification results, enhancing the system's responsiveness and flexibility.

• We evaluated FaaSMT and performed a performance comparison with Fusionize. Security analysis indicates that FaaSMT has significant advantages in monitoring and verifying task call chains while maintaining stability in performance during the deployment of serverless applications.

\section{Background and Motivation}

\subsection {Threat Model}
\label{Threat}

This work examines attacks targeting serverless applications deployed on public cloud platforms (e.g., Amazon Lambda). We assume the cloud provider's infrastructure is reliable, ensuring proper function deployment and no collusion with malicious actors. These public serverless computing platforms allow users to develop complex applications and charge based on the number of function invocations. The provider guarantees mutual isolation between customer environments and offers customizable security solutions (e.g., firewalls and virtual networks) to enhance user security. However, the untrustworthiness of functions arises from various factors, including the potential insecurity of the programming languages used by developers and possible configuration errors. For instance, attackers may manipulate the function call chain, alter the data transmission process, or exploit legitimate API interfaces to bypass security controls, leading to data breaches. Fig.~\ref{Background} illustrates the process through which users, including both legitimate and malicious users, initiate requests via an API gateway in a serverless platform \cite{shafiei2022serverless}. Malicious users may exploit vulnerabilities in the system along the attack path (\ding{172}-\ding{175}) to carry out attacks, such as bypassing authentication for business logic tampering or launching DoW attacks. These attacks can infiltrate blind spots in logging and monitoring, severely impacting the integrity of the system. To address these security threats, this study designs the FaaSMT system to ensure the secure execution of FaaS functions. FaaSMT aids users in mitigating these security threats and effectively completing requests along the path (\ding{172}-\ding{176}). The system assumes that all communications, including those between the system and the platform as well as between the platform and developers, are conducted over secure connections (e.g., HTTPS) to ensure the security of data during transmission.

\subsection {Background}

\begin{figure}[!t]
\centering
\includegraphics[width=3.5in]{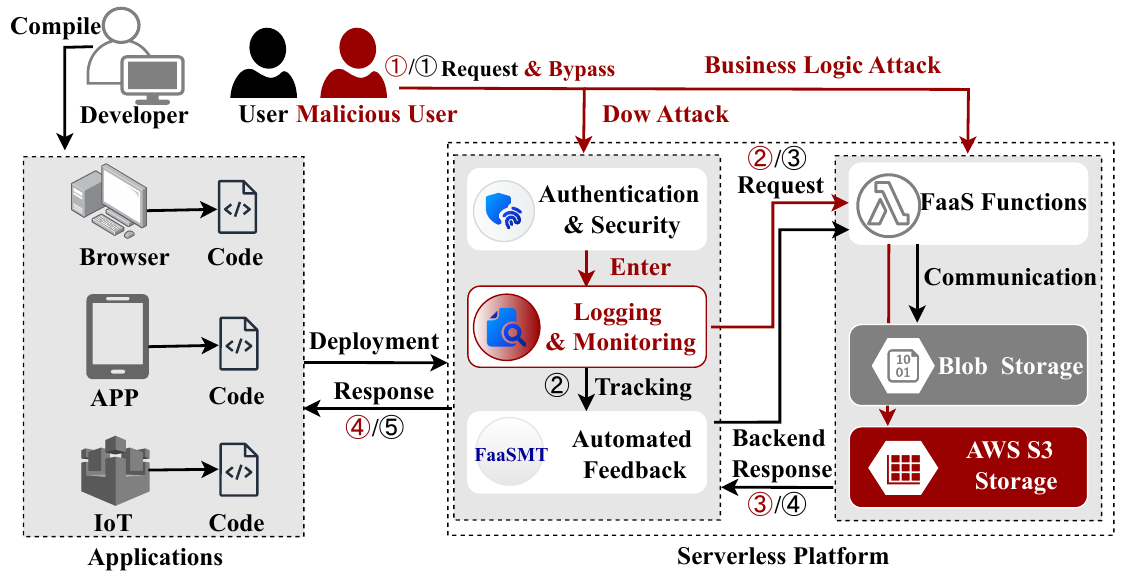}
\caption{Serverless architecture interaction processes and deployment in various application scenarios}\label{Background}
\end{figure}

\textbf{Serverless Computing:} Fig.~\ref{Background} provides an illustration of the interaction processes and deployment of a serverless architecture across a variety of application scenarios. Serverless computing, as a significant evolution in cloud computing, greatly simplifies the development and deployment of applications. Compared to traditional Infrastructure as a Service (IaaS) \cite{faragardi2019grp} and Platform as a Service (PaaS) \cite{pahl2015containerization} models, serverless architecture adopts a FaaS model, breaking applications into small, independent function units. This allows for finer-grained resource management, enabling flexible responses to changing workloads and reducing resource waste. However, this flexibility also increases the complexity of communication between functions. Communicating through cloud storage services (e.g., AWS S3 or Blob Storage) can introduce a series of security risks, including data leakage and unauthorized access. Although containerization can encapsulate applications and their dependencies, ensuring user isolation and minimizing mutual impact, it still has limitations in the face of security attacks (e.g., in 5G mobile networks) \cite{xing2023hybrid,kermabon2024perfspec}. It cannot fully address threats arising from misconfigurations and vulnerabilities. Increasing research indicates that in serverless architectures, containerized isolation does not eliminate all security risks; malicious users can still exploit vulnerabilities within functions \cite{krug2019hacking,datta2020valve,datta2022alastor}. Therefore, there is an urgent need for flexible validation mechanisms that can adaptively adjust deployment components while optimizing resource allocation and ensuring correct task execution. This approach would enable real-time monitoring of requests and the ability to track and respond to abnormal behaviour, ultimately enhancing the security and efficiency of serverless architectures.

\textbf{Merkle Tree:} The application of Merkle Tree in data integrity verification has garnered widespread attention \cite{mykletun2003providing,shamendra2023trufaas,buccafurri2024mqtt} and can take the form of either binary trees or polynomial trees. In such trees, the values of the leaf nodes are typically the hash values of data blocks, while the values of the non-leaf nodes are combinations of the hash values of their child nodes. The root node, located at the top, provides authentication permissions for accessing the integrity claims of the leaf nodes. This characteristic enables Merkle Tree to verify the integrity of function executions efficiently. During verification, it is sufficient to check the hash value of the root node to confirm that all child nodes remain untampered. This paper implements trust verification in FaaSMT using the SHA-256 hashing algorithm. During the deployment phase, relevant function execution information is extracted from the evidence storage via logging and monitoring systems and is subsequently hashed. The generated hash values are then organized into a Merkle Tree to establish an efficient data integrity verification mechanism. The root hash value of the Merkle Tree is securely stored in the managed external component AWS S3. Furthermore, to ensure the security of the root hash value, it will be encrypted in the future using KMS to prevent unauthorized access.

\subsection {Motivation}
\textbf{Limitations of Existing Approaches:} 
\begin{itemize}
    \item \textbf{Log Anomaly Detection Technology.} Intrusion anomaly detection techniques modify operating systems or system libraries through third-party functions to monitor and detect anomalies in function calls \cite{pasquier2015camflow,zeldovich2008securing}. These mechanisms primarily involve modifications to function source code and configuration files, which may violate the principle of least privilege \cite{gopalakrishna2005efficient,puresec}. Meanwhile, machine learning algorithms can dynamically identify normal and abnormal behaviours \cite{lavi2024detection,birman2020cost}. For instance, DeepLog relies on static log data, which proves inadequate in addressing dynamic log patterns. Deep learning models require continuous updates of training samples and real-time feedback to adapt to newly emerging log patterns. This not only increases hardware resource consumption but may also cause response delays, thereby impacting overall security. Consequently, these technologies still fall short of meeting the demands for distributed tracking and verification mechanisms against attacks in serverless platforms. Our research concentrates on these key issues.
    \item \textbf{Limitations of Integrity Frameworks.} Control Flow Integrity (CFI) \cite{abadi2005erlingsson} mechanisms ensure that programs execute along predefined legitimate paths by monitoring and verifying the control flow in real-time, thereby preventing the insertion and execution of malicious code. However, CFI primarily focuses on protecting the confidentiality of data and does not specifically address data integrity issues. This is where the Kalium \cite{jegan2023guarding} framework comes into play, aiming to enforce data integrity protection in serverless applications through encryption offloading techniques. Although tools for intercepting encrypted traffic may carry risks of undetected threats when implementing control flow protection, and decrypting encrypted traffic significantly increases system response times, Kalium remains committed to effectively validating function integrity in dynamic environments while maintaining system stability. This paper aims to address.
     \item \textbf{Neglected Security Risks in Resource Optimization.} Some tools designed to enhance the performance and cost-effectiveness of applications may not fully address potential security vulnerabilities \cite{schirmer2022fusionize,lin2020modeling,lin2022fine}. While these tools can adapt the deployment of components in response to fluctuations in load, attackers might strategically modify their methods, making it difficult to detect performance changes associated with covert attacks. In addition, Epsagon \cite{epsagon_aws}, a serverless monitoring tool, excels in tracing and troubleshooting; however, its graphical structure complicates the differentiation between malicious and normal traffic, especially during high-load conditions. This situation enables attackers to adjust their attack frequency and intensity to evade detection. Moreover, vulnerabilities within the applications themselves can be exploited by attackers \cite{o2020serverless,kim2022vulnerabilities} to bypass performance optimization mechanisms, leading to unauthorized data access or destruction. This difficulty in rapidly adapting to attackers' behavioural patterns poses another significant challenge that our research aims to address.
\end{itemize}
\textbf{Challenges:} Existing solutions exhibit significant limitations at various levels. To address the threat model discussed in Section \ref{Threat}, it is essential to extract and detect the execution status of applications during task execution, ensuring optimal performance in a serverless environment. Consequently, serverless frameworks based on intrusion detection face two prominent challenges:
\begin{itemize}
    \item The dynamic creation and destruction of functions typically occur within a few milliseconds, making it challenging to trace and verify the dependencies and call paths between functions. This transience not only hampers the effective implementation of existing integrity verification mechanisms but also results in inadequate monitoring of cross-function communication. In the absence of effective oversight, attackers may bypass security measures by forging function calls, tampering with data, or manipulating function behaviour. Therefore, establishing a robust integrity verification mechanism capable of effectively tracking and validating cross-function communication is of paramount importance.
    \item The short lifecycle of functions and their frequent starting and stopping necessitate that auditing and monitoring tools often rely on third-party storage for continuous activity monitoring. However, as monitoring activities increase, resource consumption rises, potentially degrading system performance. Continuous data collection introduces additional latency and computational overhead, which may also impact privacy. Thus, a key consideration for achieving efficient intrusion detection is how to adaptively adjust resources to address potential security threats without sacrificing performance, while ensuring real-time responsiveness and accuracy.
\end{itemize}
\textbf{Our Approach:} Despite the difficulties encountered when attempting to verify the accuracy and functionality of serverless applications, the fundamental design patterns intrinsic to the serverless architectural paradigm can be employed to effectively address these issues. First, the stateless nature of serverless architecture reduces the potential attack surface. To prevent data loss resulting from stateless functions, tenants should store data in external services to ensure data persistence and support monitoring and anomaly detection. Second, input-dependent functions utilize the output of one function as the input to another. The serverless architecture allows complex tasks to be decomposed into simpler functions, enabling each function to be modelled independently. This independence facilitates developers in more easily adjusting and optimizing function implementations. Furthermore, through function orchestration, these independent functions can be connected in a logical sequence, forming an effective workflow. Additionally, the term \texttt{task} described in this paper refers to the functionalities created by developers, while \texttt{function} refers to the deployable unit. Each function contains a fusion group, consisting of one or more tasks executed as part of that function. These tasks can be dynamically reallocated and fused within the same function during deployment to reduce call overhead. This specific arrangement is referred to as the \texttt{fusion setup} and tasks can also be passed to other functions to optimize resource allocation.

\begin{figure}[!b]
  \begin{center}
  \includegraphics[width=3.5in]{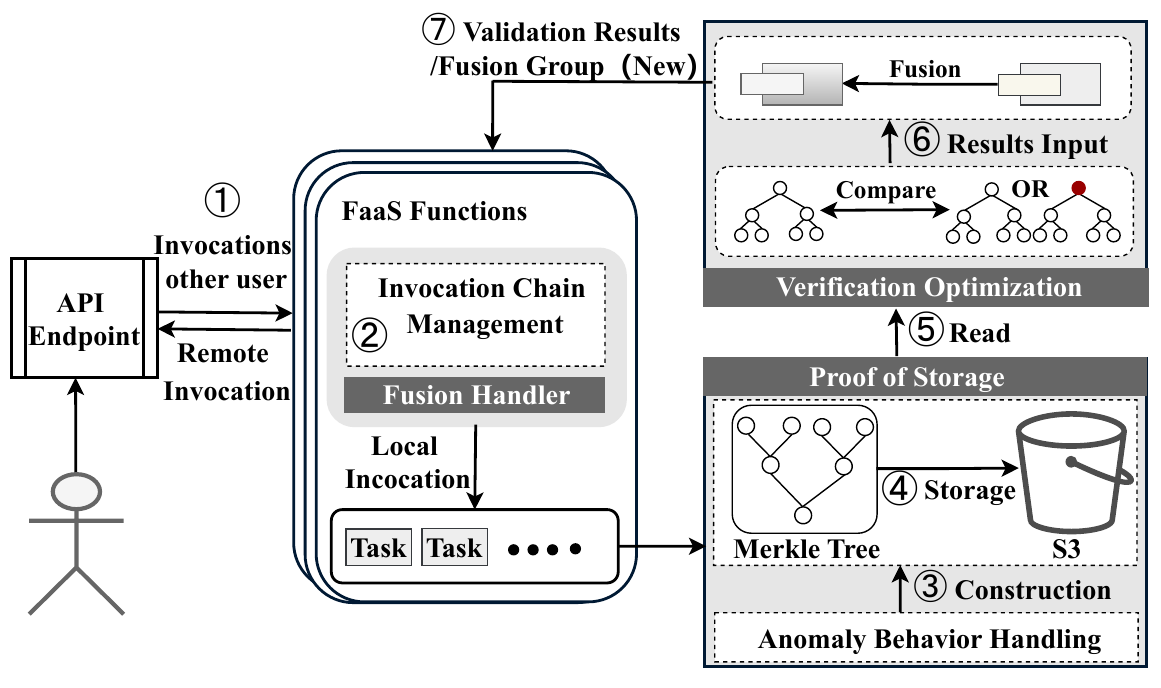}
  \caption{The Workflow of FaaSMT.}\label{FaaSMT}
  \end{center}
\end{figure}

In light of the aforementioned discussion, we proposed FaaSMT, a validation feedback-driven autonomous deployment system. It leverages the automatic monitoring capabilities of cloud FaaS platforms to verify the integrity of task-oriented applications and autonomously configures remote FaaS function calls, a process known as function fusion. For both developers and the FaaS platform, FaaSMT operates with minimal transparency. From the developer's perspective, FaaSMT acts as a driver for monitoring data and verification feedback; from the platform's perspective, FaaSMT operates on behalf of the developer, ensuring that only verified and trusted data is used during application redeployment by tracking monitoring and periodically checking data changes. As illustrated in Fig.~\ref{FaaSMT}, our approach consists of three main components: Fusion Handler, Proof of Storage, and Verification Optimizer. Next, we will outline how these components work together.

\textbf{The workflow of the automated validation feedback-driven system is as follows:} When functions are deployed on the platform, the request handler within the function container begins to listen for incoming requests, and the Fusion Handler simultaneously activates (Step \ding{172}). It retrieves the initial setup of the fusion group to which the calling function belongs from the environment variables and executes the call chain management logic, internally checking for the existence of a unique identifier for that call chain (Step \ding{173}). The final fusion setup of the fusion group determines whether to use local or remote calls. The Proof of Storage component also aggregates structured event data required for running functions from platform service logs, such as execution duration, memory usage, and additional task information for each task call. Based on this monitoring data, it performs anomaly detection on the structured event data (Step \ding{174}). To ensure the accuracy of trusted proofs, Proof of Storage constructs the Merkle Tree solely based on the processed function execution information, excluding any anomalous data, thereby generating reliable evidence stored in S3 (Step \ding{175}). The Verification Optimizer infers the application's call graph by reading the stored data. It uses scheduled tasks to retrieve all function call information stored in S3. The Verification Optimizer employs a scalable verification policy module to verify the correctness of this call graph and outputs the results (Step \ding{176}). Subsequently, based on the verification outcomes, it annotates different execution information (e.g., latency values) (Step \ding{177}). The next step utilizes a scalable optimization strategy module to derive and deploy an improved fusion setup (Step \ding{178}).

\section{Design of FaaSMT}

In this section, we propose a set of verification rules that enable FaaSMT to handle untrusted function call information. The focus is on the automated verification feedback of the FaaSMT system to ensure the integrity of the execution results of serverless applications, while improvements to other security defence strategies will be left for future research.

\subsection {Fusion Handler}
\label{Fusion}

The component is responsible for managing the function call chain and task distribution, with its primary function being call chain management. When a caller requests to invoke a function, the component first retrieves the initial configuration of the fusion group to which the function belongs from environment variables and checks for the existence of a traceID. If a traceID already exists, the component directly passes it to the relevant function call; if not, it generates a new traceID using Algorithm~\ref{alg:generate_traceid}. This algorithm first extracts configuration information from the fusionGroups array and then combines it with a randomly generated hexadecimal string and the processing function name to create a unique hash value using the SHA-256 hashing algorithm. This process ensures the uniqueness of each task execution, facilitating effective tracking and verification (lines \text{1-6}). After generating the traceID, the Fusion Handler utilizes this ID to distribute requests based on the optimal configuration obtained through examination and optimization, determining whether to use local or remote function calls. The decision-making structure relies on feedback from the Verification Optimizer to assess the success of the verification process. If verification fails, the component will notify the caller of the failure, indicating that the trust verification has failed; conversely, if verification is successful, a notification of success will be sent to the caller, indicating that the trust verification has succeeded. Ultimately, the Fusion Handler will perform task distribution according to the optimal settings obtained.
\begin{algorithm}
\small 
\caption{Generate traceID}\label{alg:generate_traceid}
\textbf{Requires:} Fusion setup $\textit{fusionGroups}$, Function handle $\textit{functionToHandle}$ \\
\textbf{Output:} Unique traceID
\begin{algorithmic}[1]
    \Function{Generatetraceid}{}
        \State $setupPart \gets \text{fusionGroups.map}(e \mapsto e.\text{join}("."))\text{.join}(",")$
        \State $randomPart \gets \text{crypto.randomBytes}(32).\text{toString}("hex")$
        \State $hashPart \gets \text{crypto.createHash}(\text{sha256}).\text{update}(setupPart + "-" + \text{functionToHandle} + "-" + randomPart).\text{digest}(\text{hex})$
        \State \Return $setupPart + "-" + \text{functionToHandle} + "-" + randomPart + "-" + hashPart$
    \EndFunction
\end{algorithmic}
\end{algorithm}

\subsection {Proof of Storage}
\label{Proof}
It is responsible for monitoring and verifying function execution information and executing the following three tasks.

\textbf{Data Monitoring and Collection.} The Proof of Storage component acquires data by executing the Fusion Handler, interacting with CloudWatch and S3 services while extracting relevant bucket and log group names from environment variables to ensure efficient AWS integration. This process enables the component to dynamically set the function's timeout based on incoming event parameters, thus accommodating the requirements of various tasks. After configuring the timeout, the component traverses all log groups to collect key information for each function invocation, including invocation time, duration, memory usage, and execution chain details, utilizing regular expressions to extract critical execution result attributes. Tracking these metrics is essential for identifying potential security threats and performance bottlenecks, as malicious activities by attackers may leave traces in the function or service logs. For instance, monitoring anomalous invocation times, memory usage patterns, and execution chain sequences can facilitate the early detection of abnormal behaviours and prompt appropriate defensive measures. Furthermore, if the FaaS platform lacks monitoring or logging capabilities, the Fusion Handler offers extended functionality to collect the necessary metrics and transmit them to a custom monitoring application. Ultimately, this data collection and transmission occur before the function invocation completes and returns results, ensuring the timeliness and integrity of the data.

\textbf{Anomaly Behavior Handling.} The method utilizes the collected monitoring data to assess the metadata of structured events to determine whether they fall within the user-defined parameters. This process ensures the rapid identification and processing of potential anomalies. Specifically, it examines the collected function call information to verify whether all structured event metadata (e.g., billing duration and memory usage) falls within predefined thresholds, and analyzes whether the invocation tasks meet the requirements of the user-defined logical business calls. If no anomalies are detected, the system continues smoothly to the subsequent processes (e.g., the construction of the Merkle Tree). Conversely, if anomalies are identified, the system filters out these anomalous tasks and employs the SHA-256 hashing algorithm on the preprocessed tasks to quantify the function call information, ensuring the integrity of the function call data. Looking ahead, this functionality aims to enhance the system's predictive capabilities to enable the early identification of user behaviour patterns.

\textbf{Generation of Trusted Proofs.} Following the completion of monitoring data collection and behaviour analysis, the Proof of Storage proceeds to generate trusted proofs. To ensure the reliability of the generated proofs, the Merkle Tree is constructed solely based on normal function execution information, excluding any data related to exception handling, thereby producing dependable evidence. Specifically, the construction of the Merkle Tree follows the rules outlined in Algorithm~\ref{alg:merkletree} to compute the root hash value. The input hash array consists of a fixed set of blocks $\{ \text{hash}(1), \text{hash}(2), \ldots, \text{hash}(n) \}$, where the hash values of these nodes are generated by hashing each function call's information. The system first checks if the input hash array is empty (line \text{2}). If the array is empty, it returns an empty root node, tree structure, and leaf nodes (line \text{3}). Next, the system duplicates the input hash list and initializes the tree structure (line \text{5}). If the length of the hash list is odd, the system duplicates the last hash value to ensure the list length is even (line \text{7}). Subsequently, the system enters a loop that continues until only one element remains in the hash list (line \text{10}). In each iteration, a new hash list is created, and combined hashes are computed by pairing adjacent hash values (lines \text{12-15}). Specifically, the system retrieves two adjacent hash values, computes their combined hash, and adds it to the new hash list while also recording the combined hash in the Merkle Tree (lines \text{16-17}), where non-leaf nodes are generated by hashing the concatenation of the left and right child nodes. It is noteworthy that if the data size is not a multiple of the block size, no padding is required, as the hash function can produce outputs for arbitrary-sized inputs. Additionally, if the number of blocks is odd, a virtual block must be added to the block list before executing the Merkle Tree algorithm (line \text{7}). Finally, the system returns the results containing the Merkle Tree's root node, tree structure, and leaf nodes (line \text{21}), ensuring that the generated Merkle Tree provides tamper-proof evidence for the call chain.

Subsequently, the components will save an array of function call information objects and the Merkle Tree proof in JSON format to a specified S3 bucket. The entire process concludes by returning a response object that includes the operation success status code and the saved call information, ensuring the correctness of the call chain proof.
\begin{algorithm}
\small 
\caption{Merkle Tree Construction}\label{alg:merkletree}
\textbf{Input:} Array of data blocks $\textit{hashes}$ \\
\textbf{Output:} Merkle Tree with $\textit{root}$, $\textit{tree}$, and $\textit{leaves}$ nodes
\begin{algorithmic}[1]
\Function{BuildMerkleTree}{$\textit{hashes}$}
    \If{$\textit{hashes} = \emptyset$}
        \State \textbf{return} $\{\text{root: } \emptyset, \text{tree: } \emptyset, \text{leaves: } \emptyset\}$
    \EndIf
    
    \State $\textit{hashList} \gets \text{copy}(\textit{hashes})$
    \State $\textit{tree} \gets \text{copy}(\textit{hashList})$
    
    \If{length of $\textit{hashList}$ is odd}
        \State $\textit{hashList}.\text{append}(\textit{hashList}[-1])$
    \EndIf
    
    \While{length of $\textit{hashList} > 1$}
        \State $\textit{newHashList} \gets \text{array}()$
        \For{$i = 0$ to $length(\textit{hashList}) - 2$ step $2$}
            \State $\textit{data1} \gets \textit{hashList}[i]$
            \State $\textit{data2} \gets \textit{hashList}[i + 1]$
            \State $\textit{combinedHash} \gets \text{calcHash}(\textit{data1} + \textit{data2})$
            \State $\textit{newHashList}.\text{append}(\textit{combinedHash})$
            \State $\textit{tree}.\text{append}(\textit{combinedHash})$
        \EndFor
        \State $\textit{hashList} \gets \textit{newHashList}$
    \EndWhile
    
    \State \textbf{return} $\{\text{root: } \textit{hashList}[0], \text{tree: } \textit{tree}, \text{leaves: } \textit{hashes}\}$
\EndFunction
\end{algorithmic}
\end{algorithm}

\subsection {Verification Optimizer}
\label{Verification}
It is the component responsible for validating and optimizing the call graph, performing two core tasks. It utilizes the array of function call information captured from the Proof of Storage along with its corresponding Merkle Tree proof to inform its validation and optimization decisions.

\textbf{Verification Strategy.} The Verification Optimizer first retrieves function call information and Merkle Tree proofs from the Proof of Storage as a reference for the verification algorithm. At the start of the verification process, this component automatically computes the hash values of all function execution information along the function call chain and reconstructs the Merkle Tree. According to the rules of Algorithm~\ref{alg: verify}, the inputs include the function call information array output from Algorithm~\ref{alg:merkletree} and the trusted proof of the Merkle Tree. During the verification process, the current block list is first duplicated, and the last element is popped off to obtain the tree information (lines \text{4-10}). If the Merkle Tree or root is empty, the status will be marked as a failure (lines \text{11-14}), and subsequent verification processing may be skipped. Then, the hash values for each block are computed and populated into a hash list (lines \text{15-19}), using the \texttt{calcHash} method to ensure that the generated hashes meet expectations. Next, a new Merkle Tree is constructed and compared with the original root (line \text{20}). If a leaf node in the proof matches the currently computed hash value, the verification process will proceed upwards through the tree to ensure the integrity of the entire call chain (lines \text{21-22}). However, if no matching leaf node is found in the Merkle Tree proof, the \texttt{findMismatch} method is called to identify the mismatched path. At this point, mismatched data blocks are retrieved from the S3 bucket and deleted (lines \text{23-29}), after which hash values are recalculated, a new Merkle Tree is constructed, and the data blocks are updated (lines \text{30-36}). Subsequently, the updated data structures are written back to the corresponding fusion group in S3 (lines \text{37-39}). Next, the integrity status of the current fusion group is recorded (lines \text{40-46}), and the overall integrity status is updated to ensure it logically reflects the results of all fusion groups (lines \text{47-50}).

\textbf{Optimization Strategy.} Based on the verification results, the Verification Optimizer initiates an iterative optimization strategy to label different execution information (e.g., latency values) for subsequent analysis. It then exports an improved fusion setup using an extensible optimization strategy module to optimize the performance of applications deployed on the cloud FaaS platform. Specifically, the iterative optimization method searches for the best fusion setup according to the grouping rules of synchronous and asynchronous functions. If none of the data blocks match, the Verification Optimizer will revert to the initial fusion setup, trigger an environment variable update, and provide this information back to the Fusion Handler. This allows the environment variables to convey the verification results and the best fusion setup to guide the request distribution in the Fusion Handler.

Furthermore, the design of the Verification Optimizer allows for the simultaneous updating and verification of historical data and the latest data during the collaborative learning verification process between the Fusion Handler and Proof of Storage. This iterative process ensures that the system can continuously optimize based on the latest validated results provided at runtime and seek the next optimal fusion setup.
\begin{algorithm}
\small 
\caption{Verify Integrity}\label{alg: verify}
\textbf{Input:} setups, deleteList \\
\textbf{Output:} integrityVerified, results, updatedDeleteList
\begin{algorithmic}[1]
\Function{VerifyIntegrity}{$setups, deleteList$}
    \State $integrityVerified \gets \text{true}$
    \State $results \gets []$
    \For{$key \in setups$}
        \State $group \gets setups[key]$
        \State $blocks \gets \text{copy}(group)$
        \State $treeInfo \gets \text{blocks.pop}()$
        \State $tree \gets treeInfo.tree$
        \State $root \gets treeInfo.root$
        \State $leaf \gets treeInfo.leaf$
        \State $status \gets \text{true}$
        \If{$\text{tree is null or root is null}$}
            \State $status \gets \text{false}$
        \Else
            \State $hashes \gets [\,]$
            \For{$block \in blocks$}
                \State $hash \gets \text{calcHash}(\text{JSON.stringify}(block))$
                \State $\text{hashes.push}(hash)$
            \EndFor
            \State $(newRoot, leafHash) \gets \text{buildMT}(hashes)$
            \If{$newRoot \neq root$}
                \State $path \gets \text{findMismatch}(leafHash, leaf, blocks)$
                \For{$idx \in \text{reverse}(path)$}
                    \State $blockToRemove \gets blocks[idx]$
                    \State $traceid \gets blockToRemove.traceid$
                    \State $deleteKey \gets key + '/' + traceid + '.json'$
                    \State $\text{S3.deleteObject}(\{
                        \text{Bucket}: bucketName,
                        \text{Key}: deleteKey
                    \}).promise()$
                    \State $\text{blocks.splice}(idx, 1)$
                \EndFor
                \State $updatedHashes \gets [\,]$
                \For{$block \in blocks$}
                    \State $hash \gets \text{calcHash}(\text{JSON.stringify}(block))$
                    \State $\text{updatedHashes.push}(hash)$
                \EndFor
                \State $newTreeInfo \gets \text{buildMT}(updatedHashes)$
                \State \textbf{Update blocks:}
                \begin{tabbing}
                    \hspace{2em} \= $\text{blocks.push}(\{$ \\
                    \hspace{2em} \= \hspace{2em} \= \text{root}: newTreeInfo.root, \\
                    \hspace{2em} \= \hspace{2em} \= \text{tree}: newTreeInfo.tree, \\
                    \hspace{2em} \= \hspace{2em} \= \text{leaf}: newTreeInfo.leaf \\
                    \hspace{2em} \= \= \text{\})}
                \end{tabbing}
                \State $fileKey \gets key + '.json'$
                \State \textbf{Update S3:}
                \begin{tabbing}
                    \hspace{2em} \= $\text{S3.putObject}(\{$ \\
                    \hspace{2em} \= \hspace{2em} \= \text{Bucket}: bucketName, \\
                    \hspace{2em} \= \hspace{2em} \= \text{Key}: fileKey, \\
                    \hspace{2em} \= \hspace{2em} \= \text{Body}: JSON.stringify(blocks), \\
                    \hspace{2em} \= \hspace{2em} \= \text{ContentType}: 'application/json' \\
                    \hspace{2em} \= \= \text{\}).promise()}
                \end{tabbing}
                \State $\text{blocks.pop}()$
                \State $deleteList[key] \gets \text{blocks}$
                \State $status \gets \text{false}$
            \EndIf
        \EndIf
        \State $group.integrityVerified \gets status$
        \State $\text{results.push}(status)$
        \State $setups[key] \gets group$
        \State $integrityVerified \gets integrityVerified \text{ and } status$
    \EndFor
    \State \Return $\{integrityVerified, results, deleteList\}$
\EndFunction
\end{algorithmic}
\end{algorithm}

\section{EVALUATION}

In our evaluation, we needed to implement a prototype of FaaSMT and Setup (Section \ref{Prototype}). Additionally, we conducted a systematic analysis of FaaSMT's performance overhead (Section \ref{Performance}) and security (Section \ref{Security}), with a particular focus on the effects of Proof of Storage and Verification Optimizer within the framework. Finally, we described the relevant limitations of the framework (Section \ref{Limitations_FaaSMT}). Specifically, we analyzed two open-source serverless applications based on Lambda using FaaSMT, evaluating their performance in terms of security and overhead. For performance assessment, we deployed FaaSMT in the AWS execution environment to measure its impact on application runtime as well as its operational overhead. One of the applications analyzed was a realistic IoT application \cite{castro2019rise,eismann2021state,grambow2021befaas,bermbach2021future}, which involves roadside sensors continuously monitoring environmental factors such as temperature, traffic volume, and air quality. The collected data is processed by AWS Lambda functions specifically tailored for each type of reading. These functions collaborate with CloudWatch for real-time analysis and manage structured and unstructured data storage through DynamoDB and S3, respectively. This application serves as an AWS-provided example for automatically updating deployment scripts following modifications to the source code or configuration.

\subsection {Prototype Implementation and Setup}
\label{Prototype}

\textbf{Prototype Implementation:} The framework presented in this paper is developed as a prototype on AWS Lambda, based on the Fusionice framework. By automatically integrating application code into multi-functional orchestrations with varying function sizes, it alleviates developers' concerns, allowing them to focus on writing application code following original programming standards without worrying about how the code is transformed into specific function implementations. Through modifications to the framework's components, we successfully integrated the enhanced algorithm, resulting in FaaSMT. The main modifications are focused on three components: Fusion Handler, Proof of Storage, and Verification Optimizer. In the prototype, the Fusion Handler is implemented with an embedded handler that routes calls within the fused tasks or externally via HTTPS to route calls across different fused groups. To prevent tampering with each call chain, we added cryptographic hashes to secure them. Additionally, Proof of Storage and Verification Optimizer are implemented as two AWS Lambda functions: the first Lambda function retrieves log data from AWS CloudWatch during task execution, analyzes this data to generate reliable Merkle Tree proofs, and stores them in AWS S3; the second Lambda function then updates the fusion setup based on the verification results. This design enables efficient and secure deployment of applications without modifying the underlying platform. However, this framework may not be suitable for applications with certain large dependencies, like large machine learning models. 

\textbf{Setup:} We evaluate the FaaSMT on an AWS EC2 instance (\texttt{t2.micro}, $1$ vCPU, $1$GB RAM, running \texttt{Amazon Linux 2} (AMI)). The instance is equipped with a $30$GB General Purpose \texttt{SSD (gp2)} EBS volume, which provides sufficient execution and data storage capacity, and offers network bandwidth of up to $5$ Gbps to support high-throughput multifunctional task orchestration. All experiments are conducted in the AWS \texttt{us-west-1} region using AWS monitoring tools to collect real-time performance data, and the framework is evaluated using test scripts. Specifically, we set the infrastructure's memory resources to $128$MB during deployment to ensure flexible adjustment of memory usage according to the varying scales of function fusion requirements.

\subsection {Performance Analysis}
\label{Performance}

\begin{figure}[ht!]
\centering
\includegraphics[width=2.5in]{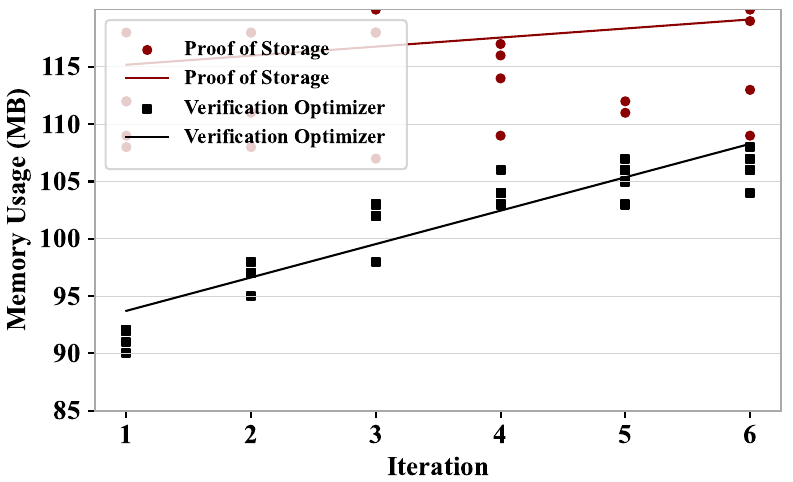}
\caption{With the increase in iterations, FaaSMT's memory usage rises significantly.}
\label{Memory}
\end{figure}

\textbf{Comparative Analysis of FaaSMT and Other Solutions Based on Merkle Tree:} As shown in Table \ref{tab:table1}, FaaSMT, PENGLAI, and TruFaaS each demonstrate unique advantages when utilizing the Merkle Tree algorithm for security checks. PENGLAI \cite{feng2021scalable} introduces a page-mapping latency overhead of $26$\%-$46$\%, but does not impact memory access bandwidth. It supports up to $1000$ concurrent enclave instances and achieves a $4$-$989$ fold improvement in startup time. TruFaaS \cite{shamendra2023trufaas}, on the other hand, exhibits a notable increase in function deployment time as the number of functions grows, primarily due to the time required for trust value insertion within the Merkle Tree's trust storage. Despite this, its performance overhead remains below $20$\%. FaaSMT's computational overhead during application runtime is illustrated in Fig.~\ref{Memory} (where the best function configuration was achieved in the sixth iteration, and will not be elaborated upon further), primarily involving Proof of Storage that records function call information to build the Merkle Tree. Although non-blocking asynchronous calls are used to alleviate the main application’s load, processing still incurs some latency. Additionally, Verification Optimize requires hash computations and the reconstruction of the Merkle Tree during the verification process, which adds extra computational overhead. Our calculations show that FaaSMT reduces the request-response latency of the example IoT application by approximately $34$\%. Given the differences in application scenarios, architectures, and testing platforms, this paper will focus solely on a direct comparative analysis with Fusionice, leaving comparisons with other solutions for future work.

\begin{table}[!t]
\caption{Performance Overhead Comparison of Different Frameworks in Merkle Tree Security Checks\label{tab:table1}}
\tiny
\begin{tabular}{|c|c|c|c|c|}
\hline
\textbf{Solution} & \textbf{Concurrency Support} & \textbf{Fast Verify} & \textbf{Platform Support} & \textbf{Performance Impact} \\
\hline
PENGLAI & \checkmark & \checkmark & RISC-V & 26\% - 46\% \\
\hline
TruFaaS & Partial & \checkmark & K8s & 10\% - 20\%  \\
\hline
FaaSMT & \checkmark & \checkmark & AWS & 34\% \\
\hline
\end{tabular}
\end{table}

\begin{figure}[ht!]
\centering
\includegraphics[width=2.5in]{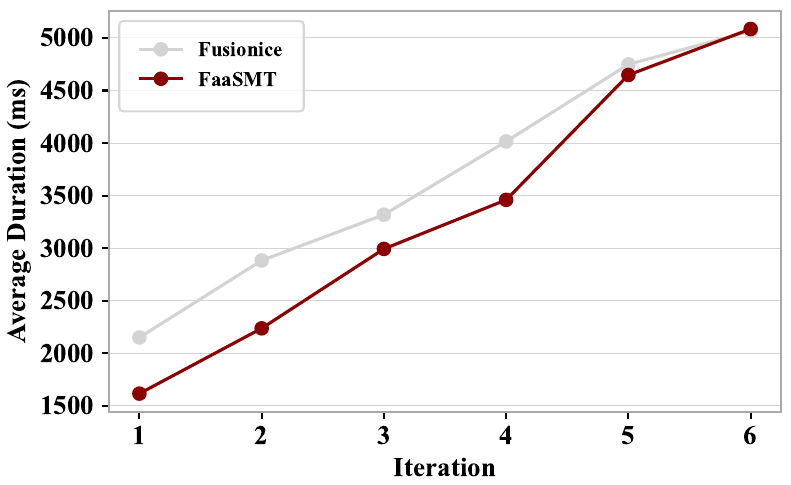}
\caption{A comparative analysis of function execution time between FaaSMT and Fusionice under different concurrent requests.}
\label{performance}
\end{figure}

\begin{figure}[ht!]
\centering
\includegraphics[width=2.5in]{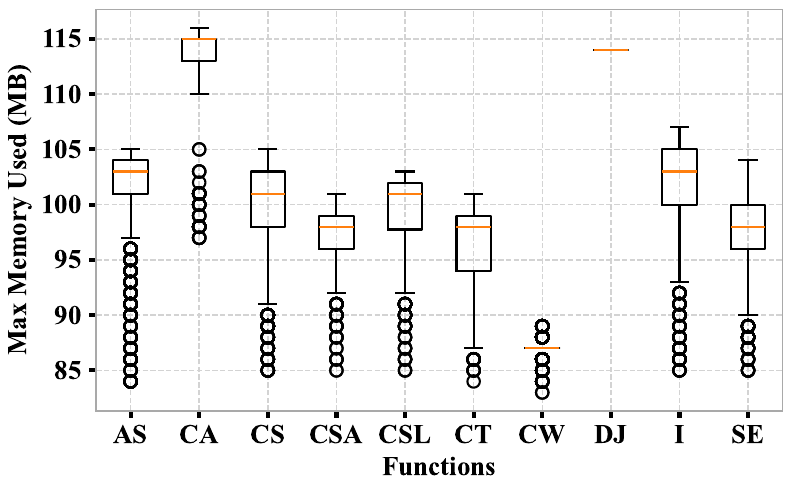}
\caption{The maximum memory usage for each function type.}
\label{maxMemory}
\end{figure}

\textbf{Comparison with Fusionice:} To evaluate the performance of Fusionice and FaaSMT, this study obtained key metrics from AWS CloudWatch, including execution duration and memory usage. During the task fusion tests conducted over seven iterations, data was collected from multiple concurrent requests ranging from $50$ to $1000$. As the number of iterations increased, the average execution time rose significantly, as shown in Fig.~\ref{performance}. The results indicate that, despite the performance overhead introduced by FaaSMT's verification mechanism, its overall performance surpasses that of Fusionice. In terms of memory usage, both frameworks increased during the iterations but did not exceed the $128$MB limit, as illustrated in Fig.~\ref{maxMemory}. This suggests that the infrastructure can effectively support varying scales of computational function fusion demands. To gain a deeper understanding of the memory usage at the end of successful runs, we calculated the total memory usage percentages for both FaaSMT and Fusionice. FaaSMT's memory usage percentage was $48.31$\%, slightly lower than Fusionice's $51.69$\%, indicating superior resource management capabilities.

\begin{figure}[ht!]
\centering
\includegraphics[width=2.5in]{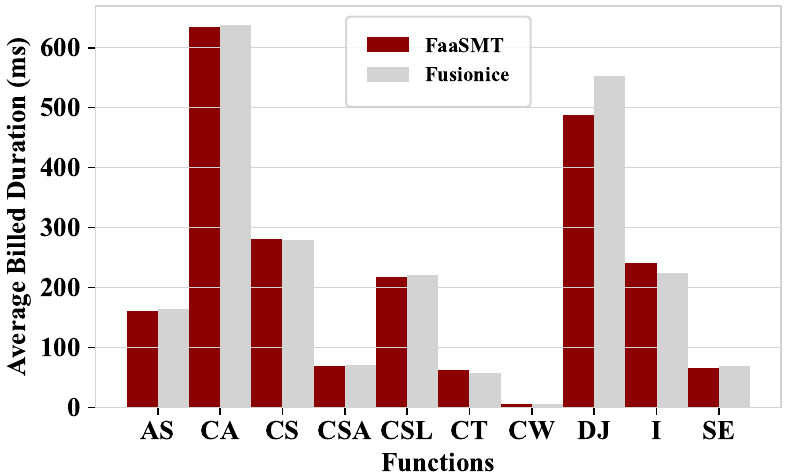}
\caption{The impact of the FaaSMT and Fusionice on IoT runtime under different loads.}
\label{standard}
\end{figure}

To evaluate the impact of the Merkle Tree algorithm on verification latency, we conducted a performance analysis of FaaSMT for each task. The key metric employed was the \texttt{billedDuration} attribute of the AWS Lambda platform, which is designed to quantify the time elapsed between the initiation and completion of a task. Each task involved multiple sets of request test cases, ranging from $1$ to $1000$, and was executed over $7$ iterations. In analyzing the function calls generated under different loads, we performed data cleaning and normalization on the raw data from FaaSMT and Fusionice to calculate the average overhead associated with building and storing function images for both frameworks. Based on this data, we obtained visual analysis results for the runtime performance in IoT applications, as shown in Fig.~\ref{standard}. The results indicate that FaaSMT has an average runtime of $11.15$\%, which is lower than Fusionice’s $11.52$\%. This finding demonstrates a significant advantage for FaaSMT in terms of execution efficiency for serverless applications.

\begin{figure}[ht!]
\centering
\includegraphics[width=2.5in]{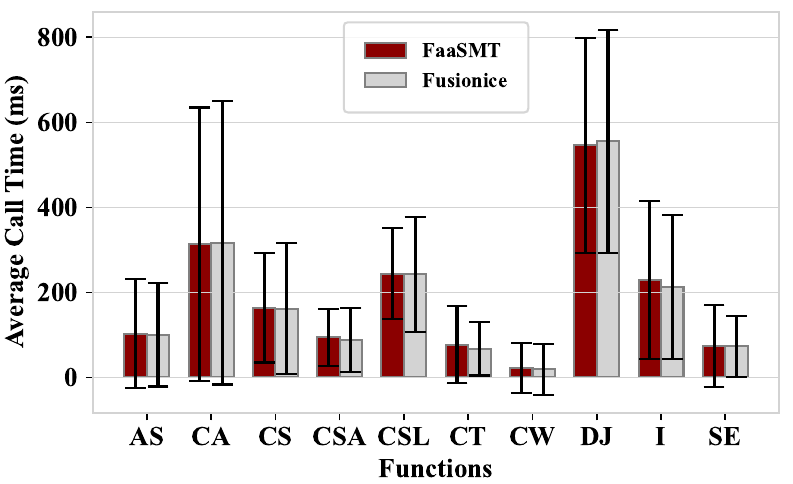}
\caption{An analysis of function call chain execution time, using the standard deviation method to measure the variability in execution time between different function calls.}
\label{standard1}
\end{figure}

We conducted a detailed analysis of the execution time of the function call chain, employing standard deviation as a means of data processing. The analysis results, as depicted in Fig.~\ref{standard1}, indicate a significant difference in the overall average call time between FaaSMT and Fusionice, with FaaSMT exhibiting an average total call time of $768899.3$ms, compared to Fusionice's $876054.7$ms. This finding suggests that FaaSMT outperforms Fusionice in execution speed during this experiment. Notably, the DJ function within FaaSMT demonstrated higher stability, evidenced by a smaller standard deviation, which further supports this conclusion.

\subsection {Security Analysis}
\label{Security}

In the context of Fusionice, the function call information provided by the FaaS platform is stored in AWS S3 services during the deployment of functions. However, attackers can exploit wallet attacks or business logic vulnerabilities to gain access to these storage locations and make modifications. This potential security risk prevents function callers from detecting whether such modifications have occurred, which may result in unexpected behaviour during function execution that differs from the original expectations set by the developers during deployment.

\textbf{DoW Attack Detection:} The DoW attack primarily targets functions that have vulnerabilities or are susceptible to exploitation. Attackers leverage code flaws, weak dependency libraries, or improper configurations within these functions to execute attacks that render services unavailable. Utilizing FaaSMT, we developed a set of test case scripts aimed at the IoT, which retrieve invocation information for all execution paths of the target application through bulk requests. Based on the threat model, we discuss the feedback mechanism of FaaSMT in automated verification to achieve more effective security responses. We conducted multiple test runs with $1000$ request calls, involving seven groups of iterative function merges. We alternated between sending normal and malicious requests in each iteration to test the system's handling capabilities for varying request frequencies. Assuming the adversary is executing a DoW attack, as illustrated in Table \ref{tab:table2}, this attack could cause the execution time of a particular function in the call chain to exceed the threshold set by AWS. During the invocation of Proof of Storage, the function call information received undergoes behavioural analysis, detecting any non-compliance with established norms. At this point, the function call chain is filtered to ensure that the Merkle Tree is ultimately constructed using legitimate invocation information, thereby generating a trusted proof. In the process of passing data to the Verification Optimizer, we assume that the attacker has managed to tamper with the transmitted data. In such cases, the Verification Optimizer conducts a second-level check and removes any tampered data to prevent contamination during the search for optimal fusion setups. By collecting the output results from the Verification Optimizer, we found that the verification results from each run of FaaSMT consistently met the expected outcomes.

\textbf{Business Logic Manipulation Detection:} In our IoT application, we also designed another attack scenario involving business logic manipulation. When applications are developed in a serverless architecture, their business logic is often decomposed into multiple single-purpose functions that interact with each other to complete tasks. Consequently, the execution order of these functions is critical for the correct execution of workflows, and any alteration in the sequence can lead to severe consequences, such as the compromise of authentication. As shown in Table \ref{tab:table2}, this highlights the general purpose of FaaSMT, which is the collection and verification of function call information. Assuming that an attacker directly manipulates the calling order of the workflow, this type of attack can be detected during the data collection phase of Proof of Storage, which identifies changes in the predefined call chain. To simulate this attack scenario, we designed test cases and conducted multiple tests for observation. In the first step, we assume that the attacker sends malicious requests, and we focus on observing the behaviour of Proof of Storage to verify its ability to filter out unexpected function call information. In the second step, we assess whether the Verification Optimizer can detect malicious behaviour during data transmission and storage, ensuring the integrity of runtime verification after constructing a trustworthy Merkle Tree and generating trusted proofs. If changes occur in either of the above two steps, we can compute the mismatched root hash based on Algorithm~\ref{alg:merkletree} and~\ref{alg: verify} and remove invalid function call information data blocks. Results indicate that during the iterative process containing malicious call chain information blocks, FaaSMT effectively identifies and removes these malicious data blocks, ensuring the trustworthiness of the computed optimal fusion setup parameters. This demonstrates FaaSMT's capability to defend against malicious tampering and unauthorized access.

\begin{table}[!t]
\caption{Comparison of Function Call Workflows and Timing for Normal and Malicious Requests\label{tab:table2}}
\centering
\scriptsize
\begin{tabular}{|c|c|}
\hline
Request Type       & Call Sequence\\
\hline
Normal Request     & I→CW(37ms)→SE(37ms)→CS(76ms)→CT(64ms)\\  &→CA(68ms) \\
\hline
DoW     & I→CW(37ms)→\textcolor{red}{SE(999999ms)}→CS(76ms)→CT(64ms)\\  &→CA(68ms) \\
\hline
Business Logic & I→CW(37ms)→SE(37ms)→CS(76ms)→\textcolor{red}{CA(68ms)}\\  &\textcolor{red}{→CT(64ms)} \\
\hline
...                & Many similar data flow calls exist                \\
\hline
\end{tabular}
\end{table}

\begin{figure}[ht!]
\centering
\includegraphics[width=2.5in]{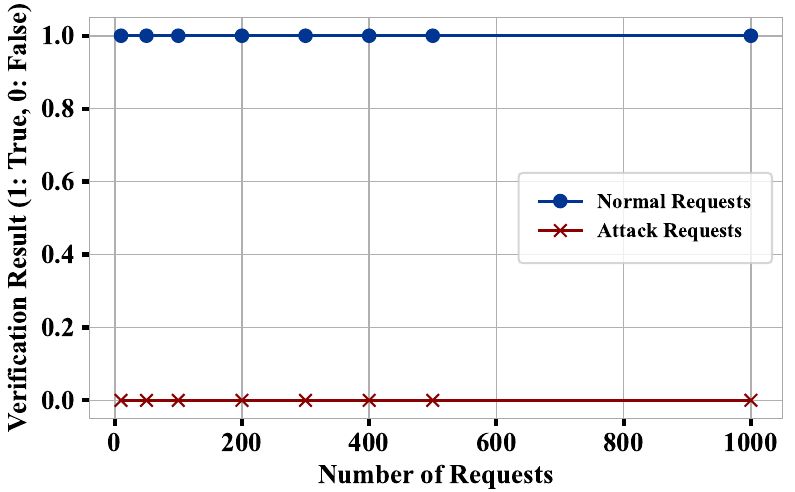}
\caption{FaaSMT Verification Outcomes Across Different Load Conditions Showing Consistent Success and Failure Rates.}
\label{result}
\end{figure}

We present an additional example of a Tree application, in which some computationally light tasks are executed synchronously, while others that are more computationally intensive are executed asynchronously. This serves to evaluate the generality of our method through the execution of multiple tests at request rates of $100$, $200$, $300$, $400$, $500$, and $1000$ requests per second. During five iterations, normal requests and malicious requests are alternately sent, thereby simulating a DoW attack scenario. Following the description in Section \ref{Proof}, we preprocess the collected logs using the Proof of Storage by handling anomalies in the properties of function executions to generate trustworthy proofs. Simultaneously, based on the description in Section \ref{Verification}, we utilize the verification results obtained through the Verification Optimize. Fig.~\ref{result} summarizes the outcomes of FaaSMT execution, demonstrating its effectiveness in verifying any suspicious function activities during each iteration.

\subsection {Limitations of the FaaSMT}
\label{Limitations_FaaSMT}

In this section, we detail the limitations of FaaSMT and directions for future research. Although FaaSMT captures rich metadata on function execution, including unusually long execution times, which aids in identifying potential abnormal behaviour, the current system may not be sufficient to differentiate between performance issues and security threats. This limitation could result in weaker attack detection capabilities. Future research could focus on enhancing data analysis methods by integrating more detailed behaviour analysis and predictive models to improve attack detection accuracy.

FaaSMT has yet to establish an effective contextual bridge with the operating system layer. During the Proof of Storage process, the framework does not account for the Intel SGX format required for Merkle Tree trust proofs. Although Intel SGX provides hardware-enhanced security, the current FaaSMT implementation does not fully leverage its isolation and encryption features, affecting the efficiency and security of trust verification with Merkle Trees. Future work should focus on integrating FaaSMT with SGX to enhance Merkle Tree trustworthiness through SGX's secure execution environment, thereby building a cross-layer trust verification system and improving overall FaaS application security. The Verification Optimizer encounters significant storage demands when handling large-scale data, and the storage requirements of the Merkle Tree grow logarithmically. Therefore, future management of storage resources requires attention. In the verification process using Merkle Tree, future implementations could utilize an incremental update method to update only the affected Merkle Tree nodes rather than recalculating the entire tree, which can significantly reduce computational burdens.

\section{Related Work}

In modern cloud computing environments, the unpredictability of serverless architectures poses challenges for existing performance monitoring, tracing, and observability tools \cite{epsagon_aws,dashbird2024,wagner2024,cordingly2020serverless}, making it difficult to comprehensively address their complexity. Some researchers have focused on tracking and recording key metrics to learn service behaviour patterns, optimizing FaaS performance while reducing costs \cite{horovitz2019faastest,wen2024joint}. These approaches enhance visibility and lay the groundwork for security response and monitoring. For instance, Alastor \cite{datta2022alastor} implemented a traceability framework through system call tracking, while other similar studies have explored how to protect information flow by improving the visibility of serverless platforms \cite{datta2020valve,wang2018peeking,mallissery2023pervasive}. These studies aim to enhance system traceability and security. Additionally, some researchers concentrate on the container level; for example, Y. Guo et al. \cite{guo2019building} applied trusted computing techniques in container environments, significantly improving security isolation. However, many current solutions have yet to achieve a comprehensive balance between performance and security. Consequently, FaaSMT has emerged as a solution in this field, focusing on encrypting and hashing source code during Terraform\footnote{\url{https://www.terraform.io/}} deployment and emphasizing trust computation for properties after function calls. To meet the demands of serverless computing, monitoring and verification tools must possess real-time inspection and anomaly detection capabilities to effectively address challenges in complex environments.

\section{Conclusion}

This paper presents FaaSMT, a trustworthy deployment framework for optimizing FaaS functions by utilizing Merkle Tree to verify functionalities and automate function fusion. Merkle Tree is employed to validate the correctness of functionalities, ensuring that each functionality remains unaltered before fusion through its hash structure. Function fusion is achieved by merging functionalities written by developers into the same function or passing functionalities to other functions as needed, thereby improving resource management. Developers can write application functionalities within a familiar function programming model, while FaaSMT automatically verifies the correctness of execution and performs function fusion. The framework is capable of handling both local and remote function calls, leveraging monitoring data to optimize the deployment and distribution of serverless applications, enhancing security while reducing latency and costs. By constructing a proof-of-concept prototype on AWS Lambda for the Node.js runtime, we validated the algorithm's significant security checking effectiveness in Tree and IoT use cases. Future work will focus on further optimizing the algorithm and integrating its optimization features at the platform level.

\bibliographystyle{IEEEtran}
\bibliography{Bibliography}

\begin{IEEEbiography}
[{\includegraphics[width=1in,height=1.25in,clip,keepaspectratio]{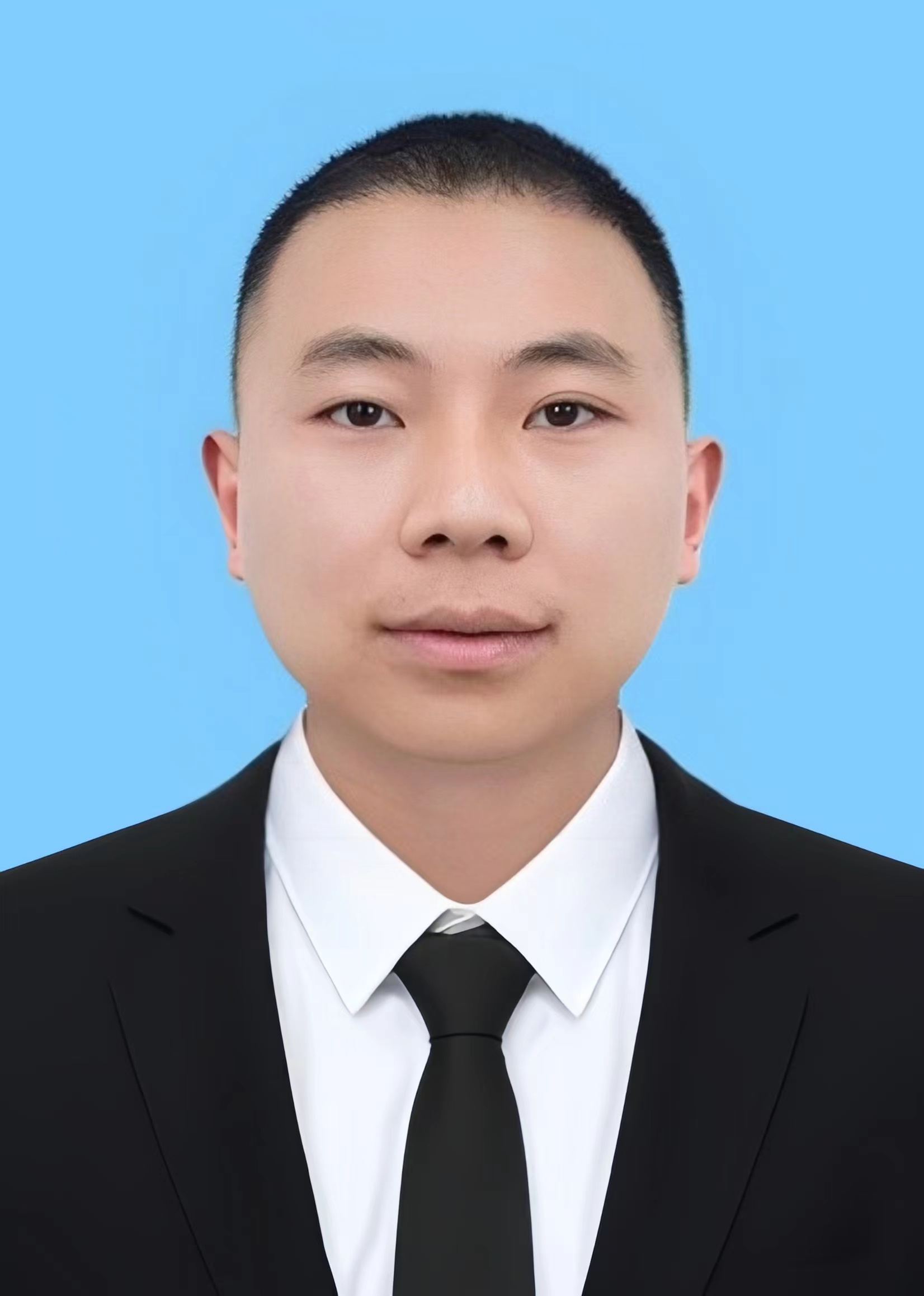}}] 
{Chuang Li}
 received the Ph.D. degree in computer science from Hunan University, Changsha, China, in 2019. He is currently an Associate Professor with the Hunan University of Technology and Business, Changsha, China. From 2017 to 2018, he was a student at NTU's Biomedical Informatics Lab, Singapore. He has authored research papers in journals such as IEEE TSC, IEEE TII, IEEE TCBB, IEEE TCE, Information Sciences, and others. His main research interests include artificial intelligence, advanced computing, and high-performance computing. 
\end{IEEEbiography}

\begin{IEEEbiography}
[{\includegraphics[width=1in,height=1.25in,clip,keepaspectratio]{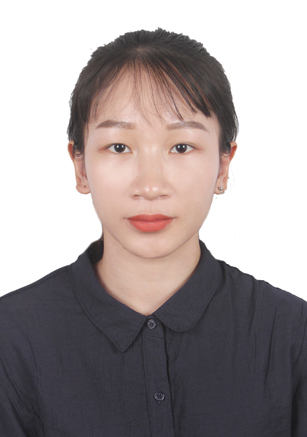}}] 
{Lanfang Huang}
 is currently pursuing the M.S. degree with the College of Computer Science, Hunan University Of Technology and Business, Hunan, China. Her research interests include cloud computing security, privacy computing, and computer system architecture. 
\end{IEEEbiography}

\begin{IEEEbiography}
[{\includegraphics[width=1in,height=1.25in,clip,keepaspectratio]{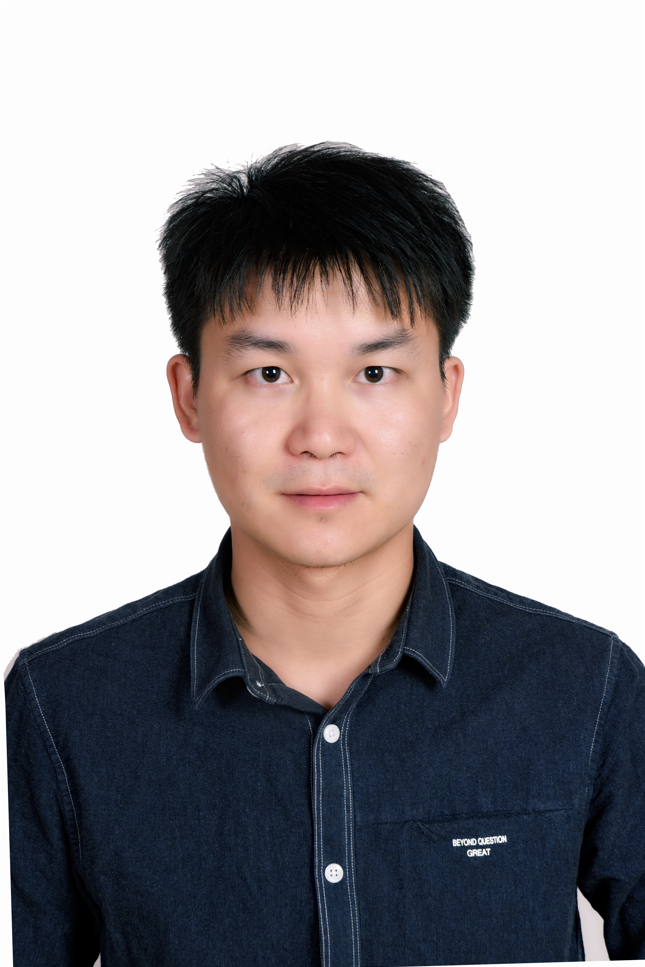}}] 
{Gang Liu}
 received the Ph.D. degree in computer science from Hunan University, China, in 2021. He was a Visiting Scholar with
the Illinois Institute of Technology from 2019 to 2020. He conducted his postdoctoral research at Hunan University from 2021 to 2023 and now holds an associate professor position at Shenzhen Institute for Advanced Study, University of Electronic Science and Technology of China. His research focuses mainly on security, high performance computing, cloud computing, and computer system architecture.
\end{IEEEbiography}

\begin{IEEEbiography}
[{\includegraphics[width=1in,height=1.25in,clip,keepaspectratio]{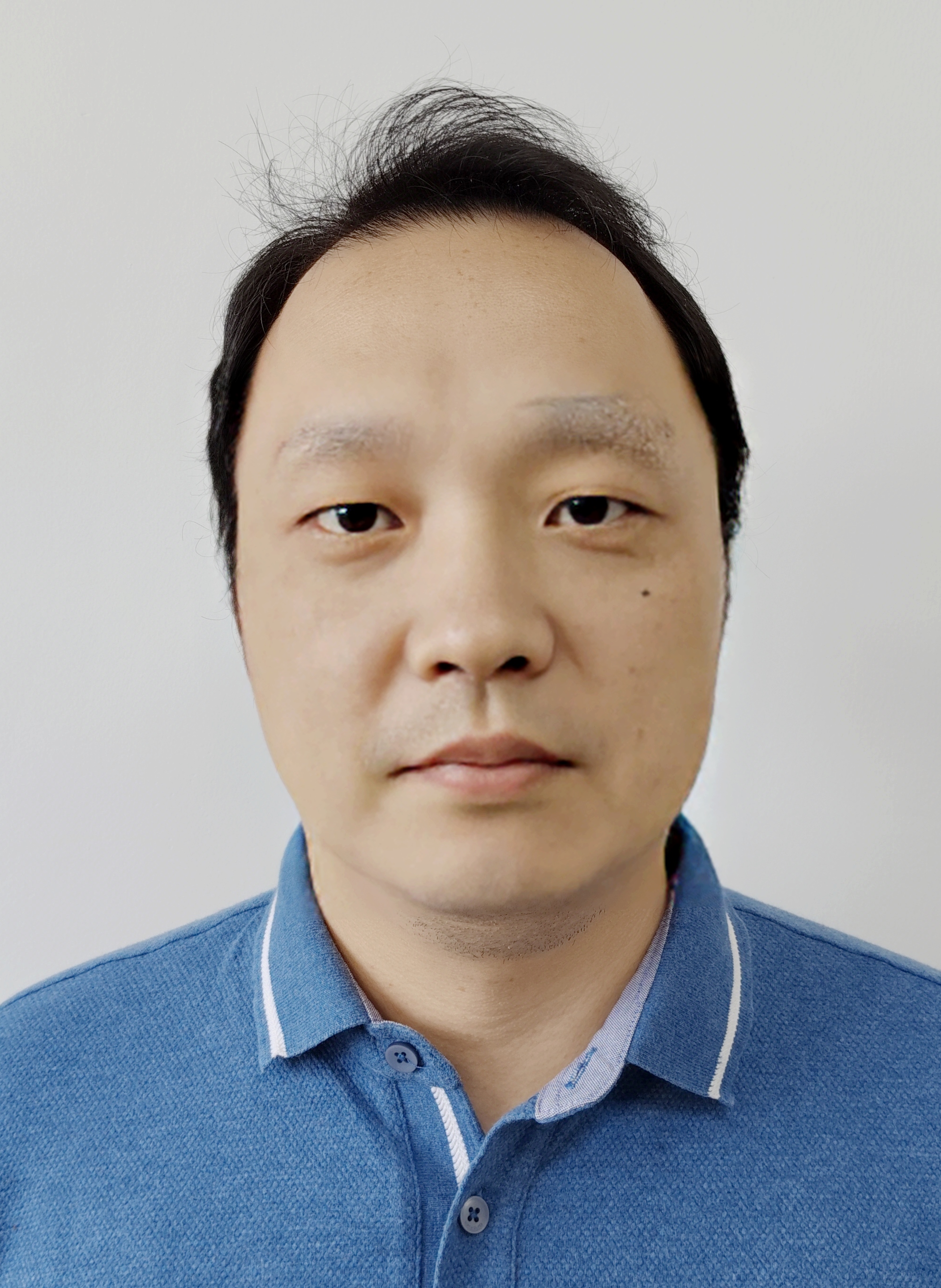}}] 
{Dian He}
 received the B.S. degree in electrical engineering and automation from Huazhong University of Science and Technology in 1999 and the M.S. degree in computer application from Central South University for Nationalities in 2005. He is currently an Associate Professor at Hunan University of Technology and Business. His research mainly focuses on machine learning and deep learning, cloud computing, and data management.
\end{IEEEbiography}

\begin{IEEEbiography}
[{\includegraphics[width=1in,height=1.25in,clip,keepaspectratio]{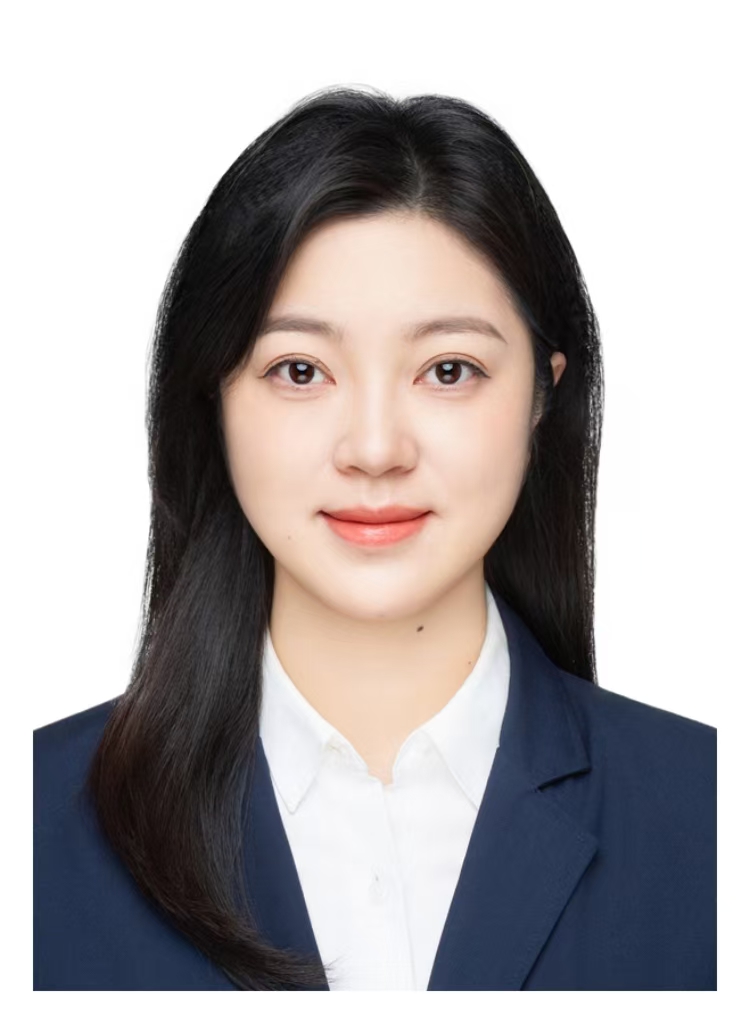}}] 
{Yanhua Wen}
 received the Ph.D. degree in electronic from the Polytech’ Nantes, Nantes, France, in 2013. She is currently a Lecturer with the Hunan University of Technology and Business, Chang Sha, China. She has produced several highly regarded scholarly works in publications including PIER, Antennas, Physics, and others. Some of her primary scientific interests include federal learning, data pricing, and artificial intelligence.
\end{IEEEbiography}

\begin{IEEEbiography}
[{\includegraphics[width=1in,height=1.25in,clip,keepaspectratio]{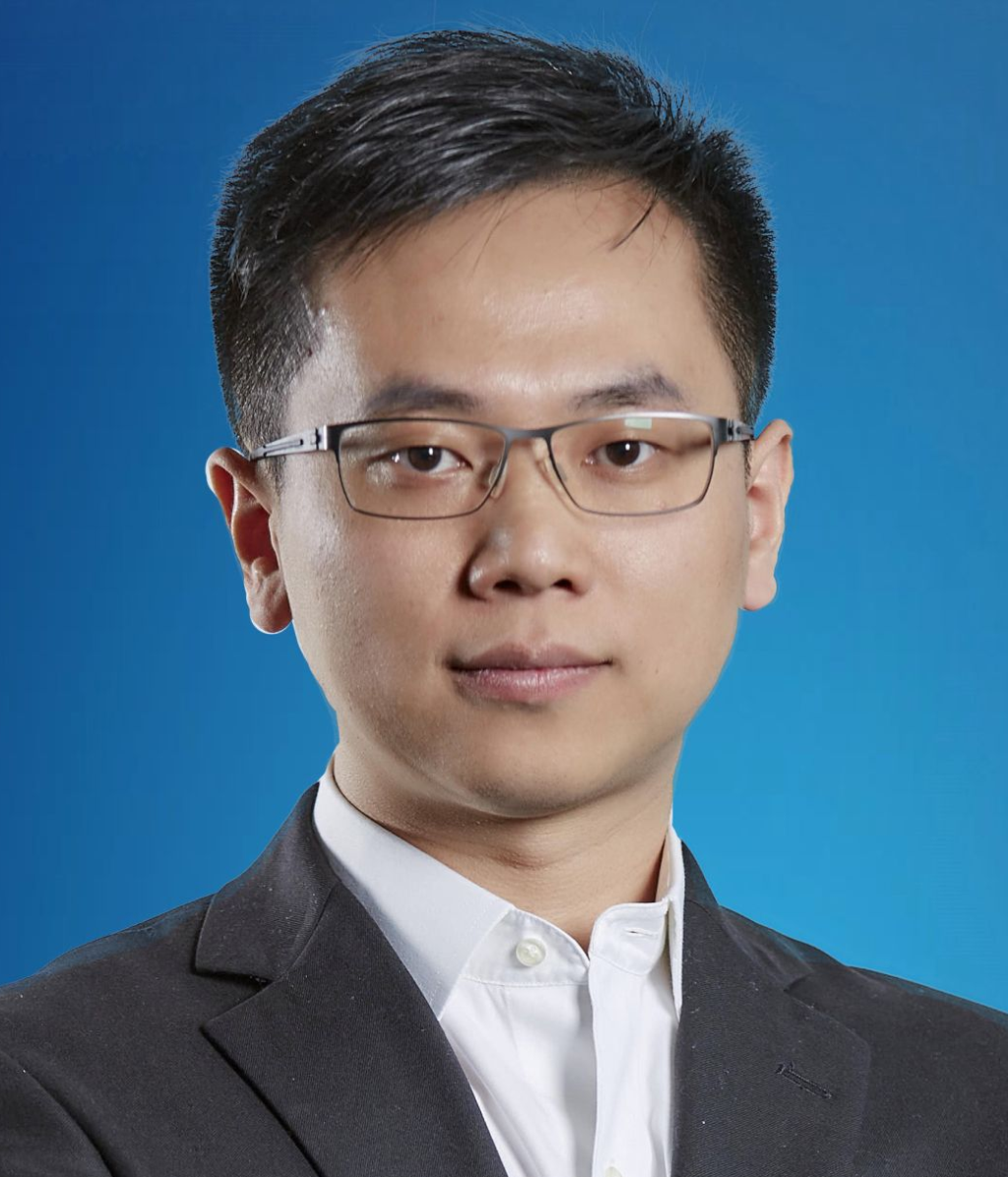}}] 
{Lixin Duan}
 received the B.E. degree from the University of Science and Technology of China in 2008 and the Ph.D. degree from Nanyang Technological University in 2012. He is currently a full professor with the School of Computer Science and Engineering, University of Electronic Science and Technology of China. His main research interests include machine learning algorithms (transfer learning, multi-modal learning, etc.) and their applications in computer vision.
\end{IEEEbiography}

\newpage

\vfill

\end{document}